\documentclass[%
 aps,
 prl,%
 amsmath,amssymb,superscriptaddress,
reprint,%
]{revtex4-1}

\usepackage{graphicx}
\usepackage{bm}

\begin{document}

\title{ Phonon Interference and Thermal Conductance Reduction in Atomic-Scale Metamaterials}

\author{Haoxue Han}
\affiliation{CNRS, UPR 288 Laboratoire d'Energ\'etique Mol\'eculaire et Macroscopique, Combustion (EM2C), Grande Voie des Vignes, 92295 Ch\^atenay-Malabry, France}
\affiliation{Ecole Centrale Paris, Grande Voie des Vignes, 92295 Ch\^atenay-Malabry, France}%

\author{Lyudmila G. Potyomina}
\affiliation {Department of Physics and Technology, National
Technical University ``Kharkiv Polytechnic Institute", Frunze str.
21, Kharkiv 61002, Ukraine}

\author{Alexandre A. Darinskii}
\affiliation {
Institute of Crystallography, Russian Academy of Sciences, Leninskii av. 59, Moscow 119333,
Russia}

\author{Sebastian Volz}
\affiliation{CNRS, UPR 288 Laboratoire d'Energ\'etique Mol\'eculaire et Macroscopique, Combustion (EM2C), Grande Voie des Vignes, 92295 Ch\^atenay-Malabry, France}
\affiliation{Ecole Centrale Paris, Grande Voie des Vignes, 92295 Ch\^atenay-Malabry, France}%

\author{Yuriy A. Kosevich}
\email[]{Corresponding author.\\
  yukosevich@gmail.com}
\affiliation{CNRS, UPR 288 Laboratoire d'Energ\'etique Mol\'eculaire et Macroscopique, Combustion (EM2C), Grande Voie des Vignes, 92295 Ch\^atenay-Malabry, France}
\affiliation{Ecole Centrale Paris, Grande Voie des Vignes, 92295 Ch\^atenay-Malabry, France}%
\affiliation{Semenov Institute of Chemical Physics, Russian Academy of Sciences, Kosygin str. 4, Moscow 119991, Russia
}%


\begin{abstract}
We introduce and model a three-dimensional (3D) atomic-scale phononic metamaterial producing two-path phonon interference
antiresonances to control  the heat flux spectrum.
We show that a crystal plane partially embedded with defect-atom arrays can completely reflect phonons
 at the frequency prescribed by masses and interaction forces.
We emphasize the predominant role of the second phonon path and destructive interference in the origin of the total phonon reflection and thermal conductance reduction
in comparison with the Fano-resonance concept.
 The random defect distribution in the plane and
the anharmonicity of atom bonds do not deteriorate the antiresonance.
The width of antiresonance dip can provide a measure of the coherence length of the phonon wave packet.
All our conclusions are confirmed both by analytical studies of the equivalent quasi-1D lattice
models and by numerical molecular dynamics (MD) simulations of realistic 3D lattices.

\end{abstract}
\maketitle
  Two-photon interference can result in a total cancellation of the photon output because of the coalescence
of the two single photons, which was first observed by Hong \emph{et al}.\cite{hong1987}. This interference
effect occurs because two possible photon paths interfere destructively, which produces the famous ``Hong-Ou-Mandel
(HOM) dip" in the detection probability of the output photons.
The HOM dip has since then been demonstrated in both the optical\cite{santori2002, beugnon2006} and
microwave\cite{lang2013} regime.
  In particular, two-photon destructive interference was recently demonstrated in a 3D optical
metamaterial\cite{wang2012}.
  Similar destructive interference effect which results in a total reflection
can be also realized in a phonon system.
For the sound waves, the two-path phonon interference antiresonance was first described in
Ref.[\onlinecite{YAK1997}], where the anomalous zero-transmission and total absorption of long-wavelength
acoustic waves in a crystal with 2D
(planar) defect were related with the  $destructive$ $interference$ between the two possible phonon paths: through
the nearest-neighbor bonds
and through the non-nearest-neighbor bonds which couple directly crystal layers adjacent to the defect atomic plane.

 Constant endeavour has been devoted to the precisely control of heat conduction. Recent efforts
concentrated on reducing the thermal conductivity $\kappa$ via nanostructured materials with
superlattices\cite{prasher2009,kim2006} and embedded nanoparticles\cite{mingo2009,mingo2010,chen2013}.
Most works attributed the reduction of $\kappa$ to the decreased
phonon life time  and thus the mean free path (MFP), which belong to the particle description of heat conduction.
However, the role of destructive phonon interference in the reduction of $\kappa$ is much less understood in the wave picture of thermal transport.

  In this Letter we introduce and model a realistic 3D
atomic-scale phononic metamaterial that allows for manipulating
the flow of thermal energy. Two-path phonon interference is generated by exploiting the phonon reflection
on internal interfaces embedded with defect-atom arrays.
  The 2D planar defects force phonons to propagate through two paths:
through unperturbed (matrix) and perturbed (defect) interatomic bonds\cite{YAK1997,YAK2008}.
The resulting phonon interference
yields transmission antiresonance (zero-transmission dip) in the spectrum of short-wavelength phonons
that can be controlled by the masses, force constants and 2D concentration of the defect atoms.
  Our results show that the patterning of the defect-atom arrays can lead to a new departure
in thermal energy management\cite{maldovan2013}, offering potential applications in
thermal filters\cite{zhang2011}, thermal diodes\cite{li2012} and thermal cloaking\cite{narayana2012,xu2014,han2014}.

\begin{figure}
\includegraphics[width=5.8cm]{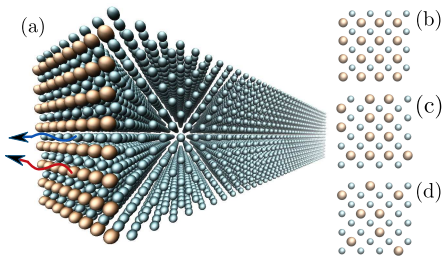}
\caption{\label{fig1} (color online).
(a) A Phononic metamaterial with a FCC lattice containing a defect plane in which an impurity-atom array
is embedded. The red and blue curve refer to the phonon path
through the impurity atom bonds and through the host atom bonds, respectively.
The presence of the two possible phonon paths can result in the ``two-path phonon interference'' antiresonance.
The brown atoms are defect atoms and the green ones are the atoms of the host lattice.
(b) The defect atoms have a periodic distribution with $f_d=50\%$.
Randomly distributed defect atoms with (c) $f_d=37.5\%$ and (d) 25$\%$ in the
defect plane.}
\end{figure}

  An atomic presentation of the 3D phononic metamaterial with a face-centered cubic
(FCC) lattice
including a 2D array of heavy defect atoms
is depicted in Fig.~\ref{fig1}(a). The defect-atom arrays are distributed periodically or randomly
in the defect crystal plane with different filling fractions $f_d$.
  When the defects do not fill entirely the defect plane, phonons have two paths to
cross such an atom array as shown in Fig.~\ref{fig1}(a),
  whereas the phonon path through the host atoms is blocked
when the defect layer is constituted by a uniform impurity-atom array, $100\%$ packed with impurity atoms.
Two types of atomic metamaterials were studied using realistic interatomic potentials:
a FCC lattice of Argon (Ar) where the defects are heavy isotopes
and a diamond lattice of Si with Germanium (Ge) atoms as the defects.
  The interactions between Ar atoms are described by the Lennard-Jones
potential\cite{juli2007}. The covalent Si:Si/Ge:Ge/Si:Ge interactions are modeled by the
Stillinger-Weber (SW) potential\cite{SW}.
  To probe the phonon transmission, MD-based phonon
wave packet (WP) method\cite{schelling2002} was used to provide the per-phonon-mode energy transmission
coefficient.
The spatial width $l$ (coherence length) of the WP is taken much larger than the wavelength
$\lambda_c$ of
the WP central frequency, corresponding to the plane-wave approximation.
All the MD simulations were performed with the {\small LAMMPS} code package\cite{LAMMPS}.

\begin{figure}
\includegraphics[width=8.0cm]{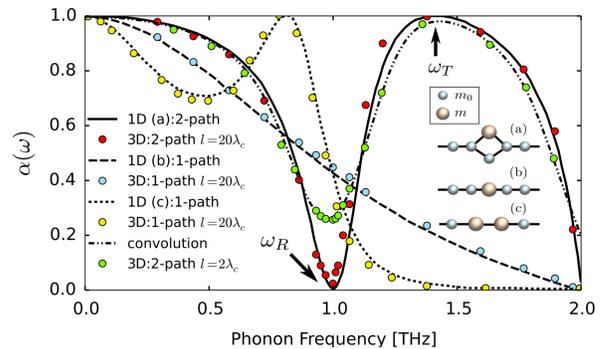}
\caption{\label{fig2} (color online).
Spectra of phonon energy transmission coefficient predicted by equivalent quasi-1D model (solid and dashed lines)
and by MD simulations (symbols) of a 3D Ar metamaterial with planar defect containing heavy
impurities, with mass $m=3m_0$.
Dashed-dotted line is the convolution of $\alpha(\omega)$ in Eq.~(\ref{eqn1}) with a gaussian WP with $l=2\lambda_c$.
Red, blue and yellow symbols represent transmission of WP with $l=20\lambda_c$ through the two paths, one path
with a single and two successive layers of defect atoms, respectively;
green symbols represent transmission of WP with $l=2\lambda_c$ through two paths.
(Inset) Three possible quasi-1D lattice models describing phonon
propagation through the lattice region containing the local defect.
Black sticks between the atoms represent atom bonds.
In the case of Ar lattice, the coefficients in Eq.~(\ref{eqn1}) are $\omega_R=1.0$, $\omega_T=1.4$,  $\omega_{max}=2.0$ and $C=0.25$.
}
\end{figure}

  The transmission coefficient $\alpha(\omega)$ of the WP with $l=20\lambda_c$,
retrieved from MD simulations of an Ar metamaterial
is presented in Fig.~\ref{fig2}.
  The incident phonons undergo a total reflection
on the defect layer at the antiresonance frequency $\omega_R$.
  Phonon transmission spectra displays an interference antiresonance profile since
the two phonon paths interfere destructively at $\omega_R$, analogous
to the two-photon interference which results in the HOM dip\cite{hong1987,wang2012}.
  A total transmission at $\omega_T$ follows the interference antiresonance,
which is reminiscent of the Fano resonances\cite{fano1961}.
  For a uniform defect-atom array, the zero-transmission antiresonance
profile will be totally suppressed and replaced by a monotonous decay of transmission with frequency.
  In the later case, only the phonon path through the defect atoms is accessible.

  We emphasize that the second phonon path is indispensable to the emergence of the zero-transmission dip,
which cannot be sufficiently described by the Fano resonance.
  We clarify this by studying the phonon transmission through
two successive uniform defect-array, where a local resonant minimum is observed instead of a zero-transmission
dip in Fig.~\ref{fig2}.
This minimum satisfies well the Fano-resonance condition\cite{fano1961} of a discrete state resonating with its continuum
background, but no zero-transmission dip occurs because of the absence of the second phonon path\cite{YAK1991,YAK1997}.
This minimum can be considered as phonon analogy of the  Fabry-P\'erot resonance in optics, which requires
only a single phonon path.
Therefore this clearly corroborates the two-path  destructive interference nature of the zero-transmission dip
in $\alpha(\omega)$.

  To further understand the phonon antiresonances caused by the interference between two phonon channels,
we use an equivalent model of monatomic quasi-1D lattice of coupled harmonic
oscillators\cite{YAK2008}, depicted
in the inset in Fig.~\ref{fig2}.
In model (a), phonons propagate through two paths: through the host atom bonds, and
through those of the impurity atoms, whereas in model (b) and (c), only the second channel
remains open.
The model (a) gives the energy transmission coefficient for
plane wave:
\begin{equation}\label{eqn1}
  \alpha(\omega)=\frac{(\omega^{2}-\omega_{R}^{2})^{2}(\omega_{\text{max}}^2-\omega^2)}
  {(\omega^{2}-\omega_{R}^{2})^{2}(\omega_{\text{max}}^2-\omega^2)+C\omega^2
  (\omega^{2}-\omega_{T}^{2})^{2}},
\end{equation}
  where $\omega_{R,T}$ are the frequencies of the reflection and transmission resonances,
$\omega_{\text{max}}$ is the maximal phonon frequency for a given polarization,
$\omega_{R}<\omega_{T}<\omega_{\text{max}}$.
$C$ is a real positive coefficient given by the atomic masses, force constants and $f_d$, $C=0$ for $f_d=0$.
  The $\omega_{R}$ frequency exists only in the presence of an additional channel which is open
for wave propagation through the bypath around the defect atom, see inset (a) in Fig.~\ref{fig2}.
  As follows from Eq.~(\ref{eqn1}) and Fig.~\ref{fig2}, $\alpha(\omega_R)=\alpha(\omega_{\text{max}})=0$ and
$\alpha(\omega_T)=\alpha(0)=1$.
  In the transmission of a narrow WP with $l=2\lambda_c$,
given by the convolution of $\alpha(\omega)$ for plane wave from Eq.~(\ref{eqn1}) with a gaussian WP in frequency domain with $l=2\lambda_c$,
the interference effect is weakened
by more frequency components when the plane-wave approximation ($l\gg\lambda_c$) is broken and the transmission at $\omega_R$ is not zero
any more, i.e. $\alpha(\omega_R)>0$, which is the case also in [1].
  As depicted in Fig.~\ref{fig2}, an excellent agreement in transmission coefficients
is demonstrated between the equivalent quasi-1D model provided by Eq.~(\ref{eqn1})
and the MD simulations of the 3D
atomic-scale phononic metamaterial with the use of realistic interatomic potential.

\begin{figure}
\includegraphics[width=8.6cm]{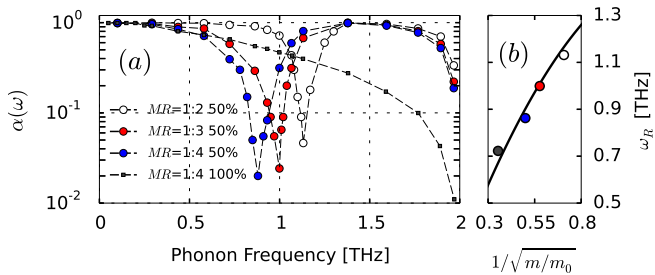}
\caption{\label{fig3}(color online).
(a) Spectra of phonon transmission coefficient $\alpha$ of longitudinal acoustic  mode in the phononic metamaterial, which
consists of 2D array of periodically alternating isotopes with different mass ratio (MR) $m/m_0$ with $f_d=50\%$ in a 3D Ar lattice.
Black squares refer to $f_d=100\%$ in the defect plane in comparison with the
$f_d=50\%$ case. Dashed lines are a guide to the eye.
(b) Isotopic shift  of the two-path phonon interference resonance
versus the mass ratio $m/m_0$. Symbols present the resonances from
MD simulations of a 3D lattice and solid line shows the analytical prediction of the equivalent quasi-1D lattice model given by Eq. (2).
}
\end{figure}

  In a lattice with atomic impurities, the substituent atoms scatter phonons due to
differences in mass and/or bond stiffness. Since no bond defect was introduced,
the loci of the resonances are only determined by the mass of the isotope atoms.
  As the defect atoms get heavier, the two-path phonon interference antiresonance becomes more
pronounced in terms of
phonon attenuation depth and width, and demonstrates a red-shift in the phonon transmission
spectrum thus impeding the long-wavelength phonons, as shown in Fig.~\ref{fig3}(a).
  The equivalent quasi-1D lattice model gives the following expression for the frequency of the transmission dip:
\begin{eqnarray}\label{eqn2}
  \omega_{R} =\omega_{\text{max}} / \sqrt{m/m_0 +1} \ ,
\end{eqnarray}
  where $m$ and $m_0$ refer to the atomic mass of the defect and host atom, $m>m_0$.
The transmission resonance at $\omega=\omega_T$ is much less sensitive to the defect mass since it is largely
determined by the mass of the host atom.
  As depicted in Fig.~\ref{fig3}(b), the spectral positions of the interference resonances
$\omega_R$ are again in an excellent agreement with the
analytical prediction of the equivalent quasi-1D lattice model given by Eq.~(\ref{eqn2}).

  In Fig.~\ref{fig3}(a), the transmission spectra for longitudinal phonons
across the uniform defect-atom array
is plotted to be compared with that of the $50\%$-filled defect-atom array.
  At the two-path interference antiresonance frequency $\omega_{R}$, an array of $50\%$ defect atoms
has a transmittance two orders of magnitude smaller than that of a uniform defect-atom array.
  The difference between the very strong phonon reflection on a $50\%$-filled defect array and
the high phonon transmission across a uniform defect array can result in
a counter-intuitive effect:
an array of segregated impurity atoms can scatter more thermal phonons than
an array with a uniform distribution of heavy isotopes.
This anomalous phonon transmission phenomenon in molecular systems can find its acoustic
conterpart in macroscopic structures\cite{YAK2008,estrada2008,liu2000}.
In Ref.[\onlinecite{estrada2008}], perforated plates
were proved to shield ultrasonic acoustic waves in water much more effectively than uniform plates.
Liu \emph{et al}.\cite{liu2000} managed to break the mass-density law for sonic transmission by embedding
high-density spheres coated with a soft material in a single layer of a stiff matrix.

\begin{figure}
\includegraphics[width=8.6cm]{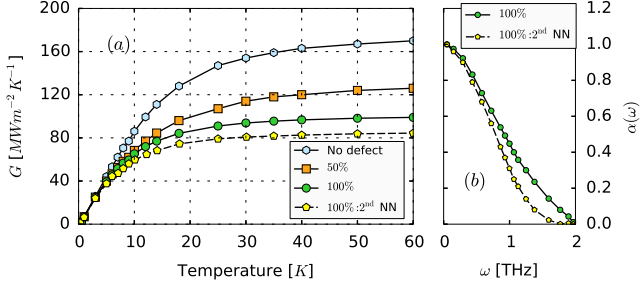}
\caption{\label{fig4}(color online).
(a) The temperature-dependent interfacial thermal conductance in a 3D Ar lattice across
a defect plane $50\%$-filled with periodic array of impurities (rectangles), a uniform defect plane with (pentagons) and without (circles) the second phonon path induced
by non-nearest-neighbor bonds,
in comparison with that of an atomic plane without defects (hexagons).
(b) $\alpha(\omega)$ for a uniform defect plane with (pentagons) and without (circles) the second phonon path.
}
\end{figure}

  We calculate the interfacial
thermal conductance $G$ by following the Landauer-like formalism\cite{khalatnikov1965}:
\begin{equation}\label{eqn3}
 G=\int\sum_{\nu}\hbar\omega(\textbf{k},\nu)
 v_{g,z}(\textbf{k},\nu)\alpha(\omega)
 \frac{\partial }{\partial T} n_{\textrm{BE}}(\omega,T)
\frac{\mathrm{d}\textbf{k}}{(2\pi)^3},
\end{equation}
where
$v_{g,z}$ is phonon group velocity in the
cross-plane direction,
$n_{\textrm{BE}}(\omega,T)=[\exp(\hbar\omega/k_BT)-1]^{-1}$ is the Bose-Einstein distribution of phonons
at temperature $T$.
The integral is carried out over the whole Brillouin zone and the sum is over the phonon branches.
  By embedding defect atoms in a monolayer, we manage to reduce the thermal conductance by $30\%$ in
respect to the case of no defects, as shown in Fig~\ref{fig4}(a). This destructive-interference-induced
effect can be used to explain the remarkable decrease of the thermal conductivity $\kappa$ of SiGe alloy
with very small amount of Ge, with respect to pristine Si\cite{garg2011}.
$G$ can be further reduced by considering the second-nearest-neighbor bonds $C_2$ between
the host atoms on the two sides of the uniform defect layer in addition to the nearest-neighbor bonds $C_1$
linking the host and adjacent defect atoms, see also Ref.[\onlinecite{YAK1997}].
This reduction comes from the suppression of phonon transmission at high frequencies, shown in Fig.~\ref{fig4}(b),
which is due to the opening of the second phonon path through the host atom bonds $C_2$
interfering destructively with the first path through the nearest-neighbor bonds $C_1$.
The emergence of the second phonon path substantially reduces $G$, by $16\%$, despite the weakness of the corresponding  bond: $C_2=0.08C_1$ in Ar lattice\cite{juli2007}.
This demonstrates another advantage of the application of the two-path destructive phonon interference for the
thermal conductivity reduction: more heat is blocked by the opening of the additional phonon paths.

\begin{figure}
\includegraphics[width=8.6cm]{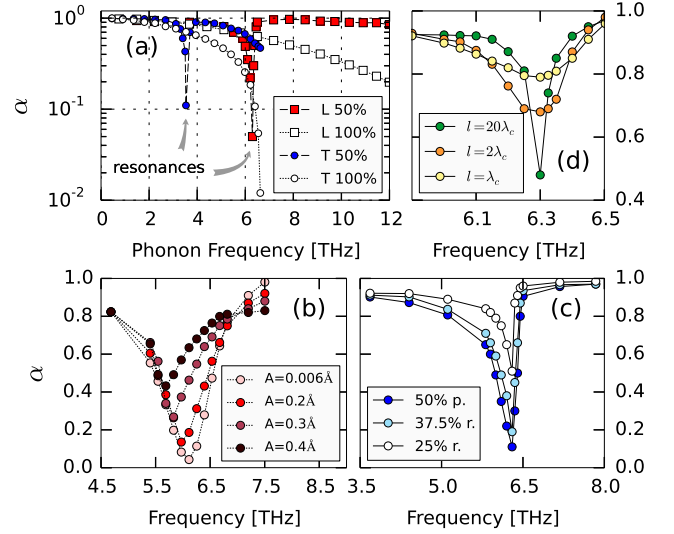}
\caption{\label{fig5} (color online).
(a) Two-path phonon interference antiresonance for both longitudinal and transverse phonons
across a segregated Ge-defect layer (red squares and blue circles) plotted along with the
non-resonant transmission across the uniform Ge-defect layer (open squares and circles)
 in a Si phononic metamaterial.
(b) Evolution of the reflection resonance versus the increasing wave amplitude.
(c) $\alpha(\omega)$ for a defect layer containing randomly dispersed Ge atoms with
$f_d=37.5\%$ and
$25\%$, compared with that of a defect layer containing
$50\%$ of periodically alternating Ge atoms. The measured $\alpha(\omega)$
were averaged over different random distributions.
(d) Resonance dip broadening in the limit of small filling fraction $f_d=5\%$
for WP with a short coherence length ($l=\lambda_c$ and $l=2\lambda_c$, yellow and orange circles), in comparison with that of almost plane-wave WP ($l=20\lambda_c$, green circles).
}
\end{figure}

In Fig.~\ref{fig5}, we report the two-path phonon antiresonance in Si crystal as the metamaterial
incorporating 2D planar impurity array of Ge atoms.
Ge and Si atom has a mass ratio of 2.57 and thus the Ge-atom array introduces both heavy
mass and bond defects due to a weaker coupling between Ge:Si than the Si:Si
interaction\cite{SW}.

  The nonlinear effects on the two-path phonon interference antiresonance was investigated by
increasing the amplitude $A$ of the incident phonon WP, as shown in Fig.~\ref{fig5}(b).
As $A$ increases, the reflection becomes less pronounced with more heat flux
passing through, which provides direct evidence of inelastic phonon scattering at the defect plane.
  The antiresonances demonstrate a red-shift in frequency due to the higher-order (cubic) terms in the
interatomic potiential.
  We also note in this connection that our computation of a quasi-1D atomic chain containing an impurity
atom characterized by non-parabolic (nonlinear) interaction potential with neighboring host atoms
agree with our MD results. The interference antiresonance remains pronounced even when the interaction
nonlinearity becomes fairly strong.
  Therefore the two-path phonon interference antiresonance in the proposed phononic metamaterial
makes it possible to control thermal
energy transport in the case of huge lattice distortions, for instance at very high temperatures.

  In contrast to light\cite{degiron2004}, even a single defect atom in a lattice plane
produces interference antiresonance for phonons because of the presence of the two phonon paths.
  Therefore, phonon reflection should be apparent even
for defect-atom arrays in the absence of periodicity because of the localized nature of the
resonance.
This is supported by further investigating the phonon
transmission through the arrays of Ge atoms, distributed in a plane in Si phononic metamaterials
with different $f_d$ and randomness.
  Strong transmission dip, similar to that in periodic arrays, remains pronounced in both cases, as shown in Fig.~\ref{fig5}(c).
This was shown experimentally to be equally valid in macroscopic acoustic metamaterials\cite{liu2000}.

Chen \emph{et al.} reduced $\kappa$ below the alloy limit by distributing random
Ge segregates in Si superlattices\cite{chen2013}.
Their \emph{ab initio} calculations showed that phonon MFP was substantially reduced
in the low frequencies\cite{chen2013}.
We note that the Ge segregates can be regarded as randomly dispersed heavy resonators which scatter low-frequency thermal phonons
by the interference antiresonances, red-shifted according to the isotopic-shift law, cf. Eq.~(\ref{eqn2}).
With this destructive interference, we can also explain the extremely low $\kappa$ found in the
In$_{0.53}$Ga$_{0.47}$As alloy randomly embedded with heavy ErAs nanoparticles\cite{kim2006}.

The decreased 2D defect filling fraction $f_d$ narrows the antiresonance width because of weakening of the relative strength of the
``defect-bond" phonon path through the crystal plane, see Fig.~\ref{fig1} and Eq.~(\ref{eqn1}).  In general, the width $\Delta \omega$ of the antiresonance dip for our two-path phonon interference is determined
by both the $f_d$ and the finite coherence length $l$ of the phonon WP.
As follows from Fig.~\ref{fig2}, for the large $f_d=50\%$, $\Delta \omega$
is not sensitive to $l$.
In the limit of small $f_d$ and for $l\gg\lambda_c$,
$\Delta \omega$ is narrow and
proportional to $f_d$, as shown in Fig.~\ref{fig5}(c) and \ref{fig5}(d).
In this limit, for the WP with short $l$, $l\sim\lambda_c$,
$\Delta \omega$ will be determined mainly by
$l$ \cite{hong1987}. From Fig.~\ref{fig5}(d), the full-width-at-half-minimum of the dip for
WP with $l=2\lambda_c$ is
$\Delta\omega=0.19$ THz. Then from the inequality
$\Delta\omega\Delta t \ge 1/2$, we get the WP width in the time domain $\Delta t = 2.6$ps and the WP
spatial width (coherence length) $l=v_g\Delta t=3.1$nm, where $v_g=1.2$km/s is the longitudinal phonon group
velocity in Si at $\omega=\omega_{R}$. This length coincides with the WP coherence length of $\approx$3.2nm, which was used
in the MD simulation shown in Fig.~\ref{fig5}(d). The width $\Delta \omega$ of the antiresonance dip for the WP with a shorter coherence length $l=\lambda_c$ is larger than that of the WP with  $l=2\lambda_c$, see Fig.~\ref{fig5}(d).
  Therefore the width of the two-path phonon interference dip in the transmission spectrum can provide a measure of
the coherence length of the phonon WP,
similar to the width of the HOM dip in the two-photon interference\cite{hong1987}.

  In conclusion, we provide comprehensive modeling of atomic-scale phononic metamaterial
 for the  control of heat transport
by exploiting two-path phonon interference antiresonances.
Thermal phonons crossing the defect plane with the two paths of different nature
undergo destructive interference, which results in the reduction of thermal conductance.
  Interference antiresonances are not deteriorated by the aperiodicity in the defect arrays and by the
anharmonicity of atom bonds.
The width of antiresonance dip provides a measure of the coherence length of the phonon wave packet.
Such patterned atomic planes can be considered as high-finesse atomic-scale phononic mirrors.
  And, at last, we would emphasize that strong optical reflections observed in 3D
stereometamaterials\cite{liu2009} can also be interpreted as photon interference
antiresonances in optically transparent plane, embedded with plasmonic nanostructures.
\\


\end{document}